\documentstyle[sprocl,feynmp]{article}

\input{epsf}
\def\Journal#1#2#3#4{{#1} {\bf #2}, #3 (#4)}

\def\NPA{{\em Nucl. Phys.} A}
\def\PRC{{\em Phys. Rev.} C}

\def\be{\begin{equation}}
\def\ee{\end{equation}}
\def\bea{\begin{eqnarray}}
\def\eea{\end{eqnarray}}

\def\LN{\Lambda N}
\begin{document}

\title{THE np $\to$ $\Lambda$p REACTION WITHIN A ONE-MESON-EXCHANGE MODEL}

\author{\underline{A. PARRE\~NO}, A. RAMOS}
\address{
Departament d'Estructura i Constituents de la Mat\`eria, Universitat de
Barcelona,\\
Diagonal 647, 08028 Barcelona, Spain}
\author{N.G. KELKAR}
\address{Nuclear Physics Division, Bhabha Atomic Research Centre, Trombay, \\
 Mumbai-400 085, India}
\author{C. BENNHOLD}
\address{
Center of Nuclear Studies, Department of Physics,
The George Washington University, Washington, DC 20052, USA}

\maketitle\abstracts{ 
With the advent of high precision proton accelerators, such as COSY in
J\"ulich, or with medium energy accelerators like RCNP at Osaka, 
it may become possible to perform a direct study of the 
strangeness changing hadronic $np \to \Lambda p$ process.
In this work we study this reaction using the one-meson-exchange weak
potential developed in Ref. \cite{PRB97} to describe the weak decay of
hypernuclei.
Both the initial $np$ and the final $\Lambda p$ states are distorted
using realistic $NN$ and $YN$ interactions.
The calculated cross sections are very sensitive to the different mesons
included in the weak transition, as well as to the strong $YN$ interaction
model that generates the distorted final state.
}

\section{Formalism}

The differential cross section per unit solid angle in the center-of-mass
system for the reaction $p n \to p \Lambda$ 
is given by the expression

\be
\frac{d {\sigma}}{d \Omega} = (2 \pi)^4 \frac{1}{s} 
\frac{|\vec p_F|}{|\vec p_I|} E_1 E_2 E_3 E_4 
\frac{1}{(2s_1+1)(2s_2+1)}
\sum_{m_{s1}} \sum_{m_{s2}} \sum_{m_{s3}} \sum_{m_{s4}}
\mid {\cal M}_{FI} \mid^2  \,\, ,
\label{eq:cs2}
\ee
with $\sqrt{s}=E_1 + E_2=E_3 + E_4$ the total available energy in the 
center-of-mass system and
${\vec p}_I$ and ${\vec p}_F$ the relative momenta of the particles
in the initial and final states respectively. 
In the distorted wave Born approximation the direct term of the
weak transition matrix elements, ${\cal M}_{FI}$, can be written as

\bea
{\cal M}_{FI}&=& \sum_{S_F M_{S_F}} \sum_{S_I M_{S_I}} \sum_{T M_T} 
\langle \frac{1}{2} m_{s_3} \frac{1}{2} m_{s_4} | S_F M_{S_F} \rangle
\langle \frac{1}{2} t_3 \frac{1}{2} t_4 | T M_T  \rangle \nonumber \\
&\times& 
\langle \frac{1}{2} m_{s_1} \frac{1}{2} m_{s_2} | S_I M_{S_I} \rangle
\langle \frac{1}{2} t_1 \frac{1}{2} t_2 | T M_T  \rangle \nonumber \\
&\times& 
\int d\Omega \int r^2 dr \,[\Psi_{\Lambda N}^{(-)}({\vec p}_F,{\vec r})]^* 
{\chi^\dagger}^{T}_{M_T} V^{w}({\vec r}\,) 
\Psi_{NN}^{(+)}({\vec p}_I,{\vec r}) \chi^{T}_{M_T} 
\eea
where $V^{w}({\vec r}\,)$ stands for the regularized weak potential 
\cite{PRB97},
which contains the exchange of pseudoscalar ($\pi,\eta$,K) and vector
($\rho,\omega$,K$^*$) mesons. The distorted wave functions, 
$\Psi_{\Lambda N}^{(-)}$ and $\Psi_{NN}^{(+)}$, 
are generated from the Lippmann-Schwinger equation
using the  
Nijmegen 93 \cite{nijmnn} or Bonn B \cite{bonn} strong $NN$ potentials 
and the Nijmegen soft-core \cite{nijmln} 
or J\"ulich \cite{HHS89} strong $YN$ ones. 

\begin{figure}[h!]
   \caption{Total cross sections for the reaction $p n \to p \Lambda$ as a
function of the proton lab momentum using the weak one-pion-exchange
potential. Dashed line: calculation omitting form factors and strong
correlations; thin solid line: including form factors and $NN$ initial
correlations; thick solid line: including form factors and both $NN$
and $\LN$ distortions. }
       \setlength{\unitlength}{1mm}
       \begin{picture}(100,100)
       \put(25,35){\epsfxsize=5cm \epsfbox{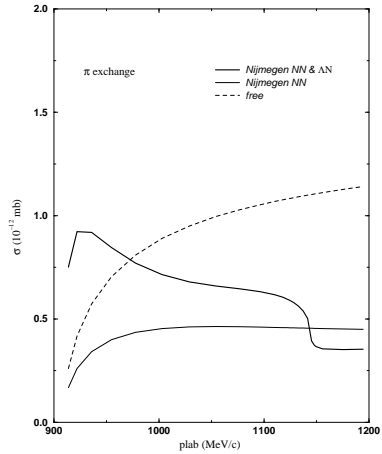}}
       \end{picture}
   \label{fig:fig1}
\end{figure}

\vspace*{-4cm}

\section{Results}

Fig. (\ref{fig:fig1}) illustrates 
the effects of the strong $NN$ and $\Lambda N$ interactions (for the
Nijmegen potentials) on the
cross sections, depicted as functions of the proton lab momentum (plab)
and when only the one-pion-exchange (OPE) mechanism is considered
in the weak transition.
When form factors and strong distorsions are omitted (``free") 
the cross
section is an increasing function of plab.
The effect of the strong repulsive part of the $NN$ interaction for this
range of momenta is to reduce 
the cross section by a factor of about 2, and is visualized by the thin
solid line. 
A substantial enhancement of the
cross section
in the low momentum region is observed when including $\Lambda N$
distorsions, due to the attractive component of the $YN$ potential in 
this region of momenta. 
As plab increases
the repulsive core of the $YN$
interaction starts having an influence in reducing the wave function
at short distances, giving rise, as a consequence, to
smaller cross sections.
The opening of the $\Sigma N$ channel, coupled to the $\Lambda N$ one
through the strong interaction,
shows up as a step in the cross section at a proton lab momentum 
around $1140$ MeV/c.

\begin{figure}[h!]
   \caption{Total cross sections for the reaction $p n \to p \Lambda$ as a
function of the proton lab momentum. The strong distortions are
generated with (a) the $NN$ Nijmegen 93 and $YN$ Nijmegen Soft Core models
(b) by the $NN$ Bonn B and $YN$ J\"ulich A.
Dashed line: $\pi$-exchange only; thin solid line: $\pi + \rho$;
thick solid line: full set of mesons.}
       \setlength{\unitlength}{1mm}
       \begin{picture}(100,100)
       \put(5,27){\epsfxsize=8cm \epsfbox{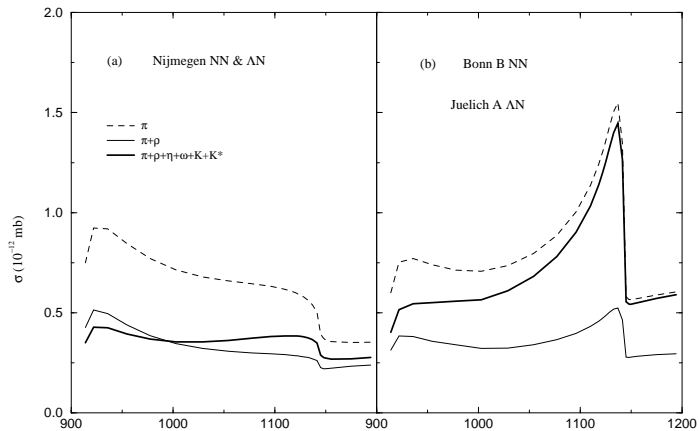}}
       \end{picture}
   \label{fig:fig2}
\end{figure}
\vspace*{-3.5cm}
The effect of adding the different mesons in the weak transition
potential is visualized in Fig. (\ref{fig:fig2}). 
The differences in the cross sections depicted in Figs. (\ref{fig:fig2}a) and
(\ref{fig:fig2}b) have to be understood as coming from the 
different $YN$ interaction models used in the calculation, since we
have checked that,
in the absence of $p
\Lambda$ distortions,
the cross sections calculated with the scattered $p n$ waves,
obtained from either the Nijmegen 93 or the Bonn B potentials, 
give practically
the same results. 
In Fig. (2b) we can see that 
already at the level of only pion-exchange
the J\"ulich A model shows a clear enhancement close
to the $p \Sigma$ threshold.

\begin{figure}[h!]
   \caption{Different partial wave contributions to the cross section
for the $p n \to p \Lambda$ reaction. Only the pion is considered in
the weak transition potential. The $NN$ and $\LN$ distorted
waves are generated with (a) the Nijmegen models and (b) the Bonn B and 
J\"ulich A models respectively.}
       \setlength{\unitlength}{1mm}
       \begin{picture}(100,100)
       \put(5,30){\epsfxsize=8cm \epsfbox{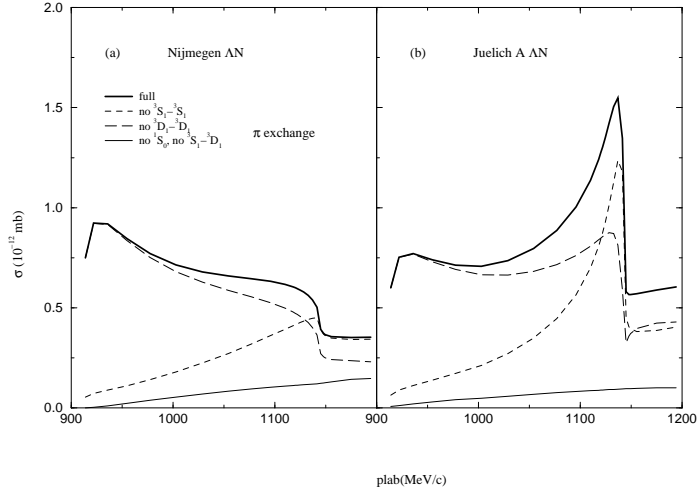}}
       \end{picture}
   \label{fig:fig3}
\end{figure}

An explanation for the differences observed between Figs. (2a) and (2b)
can be traced by analyzing Fig. (\ref{fig:fig3}), where we have plotted
the cross sections obtained by removing
the $^3S_1 \to ^3S_1$ component of the distorted $p \Lambda$ 
wave function (dashed lines), the 
$^3D_1 \to ^3D_1$ one (long-dashed lines) and all 
the \mbox{$L=0$} $\Lambda p$ partial waves for both potentials (thin solid lines).
The stronger $^3S_1-^3D_1$ channel of the J\"ulich interaction, relative
to the Nijmegen one, is the reason for the much more enhanced cross section
at the $N \Sigma$ threshold.
Suppressing the $^1S_0$ partial waves affects the cross sections negligibly
and cannot be seen in Figs. (3a) and (3b). 

\vspace*{-4cm}

\section*{References}
\begin{thebibliography}{99}

\bibitem{PRB97} A. Parre\~no, A. Ramos, and C. Bennhold, 
\Journal{\PRC}{56}{339}{1997}.

\bibitem{nijmnn} V.G. Stoks, R.A.M. Klomp, C.P.F. Terheggen and J.J. de Swart,
\Journal{\PRC}{49}{2950}{1994}.

\bibitem{bonn}R. Machleidt in 
{\em Computational Nuclear Physics 2. Nuclear Reactions},
Eds. K. Langanke, J.A. Maruhn and S.E. Koonin (Springer-Verlag, 1993).

\bibitem{nijmln}
P.M.M. Maessen, Th. A. Rijken and J.J. de
Swart, \Journal{\PRC}{40}{2226}{1989}.

\bibitem{HHS89}
B. Holzenkamp, K. Holinde, and J. Speth, 
\Journal{\NPA}{500}{485}{1989}.

\end{thebibliography}
\end{document}